\documentclass{iopart}
\usepackage{graphicx}

\begin{document}
\title[New Waveform Consistency Test for Inspiral Searches]{A New Waveform Consistency Test for Gravitational Wave Inspiral Searches}


\author{Peter Shawhan$^1$
and Evan Ochsner$^2$
}
\address{
$^1$ LIGO Laboratory, California Institute of Technology, Pasadena, CA 91125, USA\\
$^2$ Department of Physics, The University of Chicago, Chicago, IL 60637, USA
}

\eads{pshawhan@ligo.caltech.edu}


\begin{abstract}

Searches for binary inspiral signals in data collected by
interferometric gravitational wave detectors utilize matched
filtering techniques.  Although matched filtering is optimal in the
case of stationary Gaussian noise, data from real detectors often
contains ``glitches'' and episodes of excess noise which cause filter
outputs to ring strongly.  We review the standard $\chi^2$ statistic
which is used to test whether the filter output
has appropriate contributions from several different frequency
bands.  We then propose a new type of waveform consistency test
which is based on the time history of the filter output.  We apply
one such test to the data from the first LIGO science run and show
that it cleanly distinguishes between true inspiral waveforms and
large-amplitude false signals which managed to pass the standard
$\chi^2$ test.

\end{abstract}


\vspace{0.25in}

\begin{indented}
\item Submitted to \CQG for GWDAW-8 Proceedings issue
\end{indented}

\vspace{0.25in}

\section{Introduction}
\label{intro}

A binary system of two neutron stars or black holes
in a close orbit loses energy
and angular momentum through the emission of gravitational radiation,
causing the orbital distance to decrease.  In the final ``inspiral''
stage of this evolution, the emitted gravitational waves rise in
frequency and amplitude at an accelerating rate, until the orbit
becomes unstable and the objects coalesce.  The exact form of this
``chirp'' waveform depends on the masses and spins of the binary
components.  In the case of a double neutron star system, the inspiral
waveform spends many seconds within the sensitive frequency band of
the large ground-based interferometric gravitational wave detectors
which are now collecting scientific data or are being commissioned
(TAMA300, LIGO, GEO 600, and VIRGO), and spin effects are believed to
be negligible.  Thus, template waveforms for these systems can be
calculated accurately and can be used to search for this class of
signals by {\em matched filtering}~\cite{MF1,MF2}, which performs a
phase-coherent correlation of the data with the template, varying the
coalescence time parameter.  Binary systems involving low-mass black
holes (up to perhaps $\sim$$50~M_\odot$) also inspiral within the
frequency band of ground-based interferometers and can be searched for
with matched filtering, although the higher mass implies shorter
template durations (down to $\ll$$1$~s) and more severe relativistic
and spin effects, requiring searches to be done in an expanded
parameter space~\cite{BCV}.

Simple matched filtering is the optimal detection strategy if the
detector noise is stationary and white.  In the case of stationary
colored noise, optimal performance is obtained by filtering in the
frequency domain with a weight inversely proportional to the power
spectral density of the noise, as we will review briefly in
section~\ref{filtering}.  However, real gravitational wave detectors
are commonly found to suffer from {\em non-stationarity}, either in
the form of ``glitches'' (highly localized in time) or as roughly
adiabatic variation of the broadband noise level over short time
scales.  Either type of non-stationarity can strongly excite a matched
filter, leading to false ``triggers''
when the filter output amplitude exceeds a predetermined threshold.
Therefore, it has become standard to test the consistency of the
trigger-generating data with the template waveform by calculating a
$\chi^2$ statistic, defined in section~\ref{chisq}.  This value is normally
small for a real signal but tends to be large for triggers caused by
non-stationary noise.

Rejecting triggers with large $\chi^2$ eliminates many inspiral
triggers caused by non-stationary noise, but some manage to pass this
test.  This became particularly clear in follow-up examination of
inspiral event candidates found by the analysis of the first LIGO
science run (called ``S1''~\cite{S1DET}), as illustrated in
section~\ref{s1events}.  Guided by the characteristics of these event
candidates, in this paper we propose a new type of waveform
consistency test which is based on the time history of the output of
the matched filter in the vicinity of the trigger.  As described in
section~\ref{tests}, this type of test has the advantage of being
simple to implement and has negligible computational cost.  In
section~\ref{s1rerun}, we apply an empirically chosen test of this
type to the LIGO S1 data and show that it eliminates many of the
largest-amplitude triggers found by the S1 inspiral search without
reducing the efficiency for detecting real signals.  Although the
choice and tuning of a test of this type depends on the exact nature
of the non-stationarity in the detector noise, it is reasonable to
expect that waveform consistency tests of this type may lead to
significantly cleaner inspiral searches when analyzing data from other
science runs and other detectors.

\section{Review of matched filtering for inspiral searches}
\label{filtering}

We will focus on matched filtering as implemented for the search for
binary neutron star inspirals in LIGO S1 data~\cite{S1BNS}, using
similar notation.
An inspiral produces a gravitational wave strain in the
interferometer, $h(t)$, which depends on the intrinsic physical
properties of the binary system, the coalescence time $t_c$, and the
position and orientation of the system relative to the interferometer.
For low-mass systems, the masses of the two components are the only
relevant physical parameters.  The effect of the position and
orientation of the system on the received signal can be represented by
just two parameters, an {\em effective} distance $D_{\rm eff}$ and a
signal phase $\alpha$.  Thus, we may represent a template waveform as
$h^I(t-t_c;\alpha)$, where the index $I$ represents a point in the
space of intrinsic parameters and the template is normalized to
correspond to a signal with a certain effective distance.
From this point onward we will omit the
index $I$, so that the formulae will refer to any individual template
from the ``bank'' of templates which is used to cover a region of
intrinsic parameter space.

A further simplification is obtained by transforming the template to
the frequency domain using the {\em stationary phase
approximation}~\cite{SPA}, so that the template has the form
$e^{i \alpha} \tilde{h}_c(f)$ where $\tilde{h}_c(f)$ is independent of
$\alpha$.  Thus, we can simply use $\tilde{h}_c(f)$ ({\it i.e.}, the
template with $\alpha=0$) to filter the data and then consider the
{\em magnitude} of the result, effectively maximizing over the
signal phase $\alpha$ analytically.  The Wiener optimal filter, which
dictates that noisy frequencies should be suppressed, is easily
applied in the frequency domain to the Fourier-transformed data from
the detector, $\tilde{s}(f)$:
\begin{equation}
\label{e:xfilter}
  z(t) = 4 \int_{0}^{\infty}
    \frac{\tilde{h}_c(f)\tilde{s}^\ast(f)}{S_n(f)}\,e^{2\pi ift} \; df
\end{equation}
where $S_n(f)$ is the power spectral density of the detector noise,
estimated from nearby data.  The output of the matched filter is
\begin{equation}
  \rho(t) = \frac{|z(t)|}{\sigma}
\end{equation}
where
\begin{equation}
\label{e:variance}
  \sigma^2 = \frac{\left\langle |z(t)|^2 \right\rangle}{2}
  = 4 \int_{0}^{\infty} \frac{\left|\tilde{h}_c(f)\right|^2}{S_n(f)} \; df
\end{equation}
normalizes $\rho$ so that it can be interpreted as an amplitude
signal-to-noise ratio (SNR).  If the detector noise is stationary (and there
is no signal present), then $\rho(t)$ will be a random variable
with $\left\langle \rho^2 \right\rangle = 2$.

The filter output $\rho(t)$ is evaluated
at a set of discrete times $t$ with a time step shorter than
the period of the inspiral waveform at its highest
frequency.  For example, the LIGO S1 analysis used a time step of
$1/4096$~s.  A true inspiral signal in the data would lead to a narrow
peak in $\rho(t)$ at the coalescence time, reflecting the fact that
each time sample in the filter output is the appropriate coherent sum
of signal power distributed over time and frequency in the input time
series.
Accordingly, the inspiral search algorithm essentially consists of
looking for local maxima of $\rho(t)$ (separated in time by at least
the length of the template) which exceed some fixed threshold
$\rho^\ast$.  Each such maximum is called a ``trigger'', characterized
by $\rho_{\rm max}$, and is subjected to further evaluation.  The
threshold $\rho^\ast$ is chosen for practical reasons, to yield a
manageable number of triggers.

\section{The standard $\chi^2$ test}
\label{chisq}

The technique of calculating a $\chi^2$ to check the consistency of a
trigger with the expected waveform was developed several years
ago~\cite{GRASP} and has been utilized in published inspiral
searches~\cite{40m,TAMAinspiral,S1BNS}.  The inspiral template is
effectively divided into $p$ sub-templates labeled by $l = \{1,2,...,p\}$,
each of which contains a different frequency band of the original
template.  The data is filtered using each of these templates:
\begin{equation}
  z_l(t) = 4 \int_{F_{l-1}}^{F_l}
    \frac{\tilde{h}_c(f)\tilde{s}^\ast(f)}{S_n(f)}\,e^{2\pi ift} \; df
\end{equation}
where the frequency boundaries $F_l$ are chosen
so that each of the sub-templates should, in the
absence of noise, contribute equally to the total signal, {\it i.e.}\ $\langle
z_l(t_c) \rangle = z(t_c) / p$.  Using these values, the $\chi^2$
statistic is calculated as
\begin{equation}
  \chi^2(t) = \frac{p}{\sigma^{2}}\sum_{l=1}^p \left|z_l(t)-z(t)/p\right|^2
\end{equation}
evaluated at the inferred coalescence time of the original trigger.
Note that this definition demands that the sub-templates be consistent
with the full template in both amplitude and phase.  If the signal in
the data matches the template exactly, then this $\chi^2$ statistic
will follow a chi-squared distribution with $2p-2$ degrees of freedom.

\begin{figure}[tb]
  \begin{center}
  \includegraphics[width=2.5in]{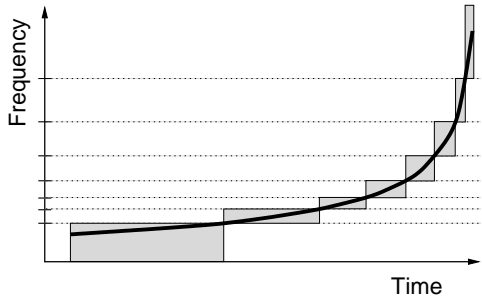}
  \end{center}
  \caption{Graphical representation of the $\chi^2$ statistic in a
  time-frequency plane.  The thick black line shows the frequency of
  the chirp waveform rising with time.  The division of the waveform
  into $p$ frequency bands ($p=8$ in this case, with boundaries indicated by
  the dotted horizontal lines) essentially creates $p$ sub-templates, each
  of which uses the data in that frequency band over a limited time
  interval, represented by the shaded boxes.}
  \label{fig:slice}
\end{figure}

Figure~\ref{fig:slice} is a conceptual illustration of how the data
contributes to the $\chi^2$ statistic.  Because the chirp frequency rises
monotonically, dividing the waveform into frequency bands is
essentially equivalent to dividing it into time intervals.  Each
partial filter output $z_l(t_c)$ is affected only by the data in its
frequency band and time interval (the shaded boxes in the figure),
with the appropriate time delay to relate it to the output of the full
filter at the inferred coalescence time.  Outside of this ``chain'' of
boxes following the chirp, no other regions of the time-frequency
plane affect the value of the $\chi^2$ statistic.

Triggers with $\chi^2$ values above some threshold are discarded.
However, for large-amplitude signals, the $\chi^2$ statistic is highly
sensitive to any small mismatch between the waveform and the template
used for filtering, which is unavoidable given the discrete template
bank used to perform the search.  Therefore, a {\em variable} $\chi^2$
threshold, with an appropriate
dependence on $\rho_{\rm max}$, is needed to avoid rejecting true
inspiral signals which have very large amplitudes.  For example, the
LIGO S1 inspiral analysis~\cite{S1BNS} required $\chi^2$ to satisfy
\begin{equation}
\chi^2 < 40 + 0.15\;\rho_{\rm max}^2 \; .
\label{eq:modchisq}
\end{equation}

\section{Inspiral event candidates from the LIGO S1 science run}
\label{s1events}

\begin{figure}[tb]
  \begin{center}
  \includegraphics[width=5.0in]{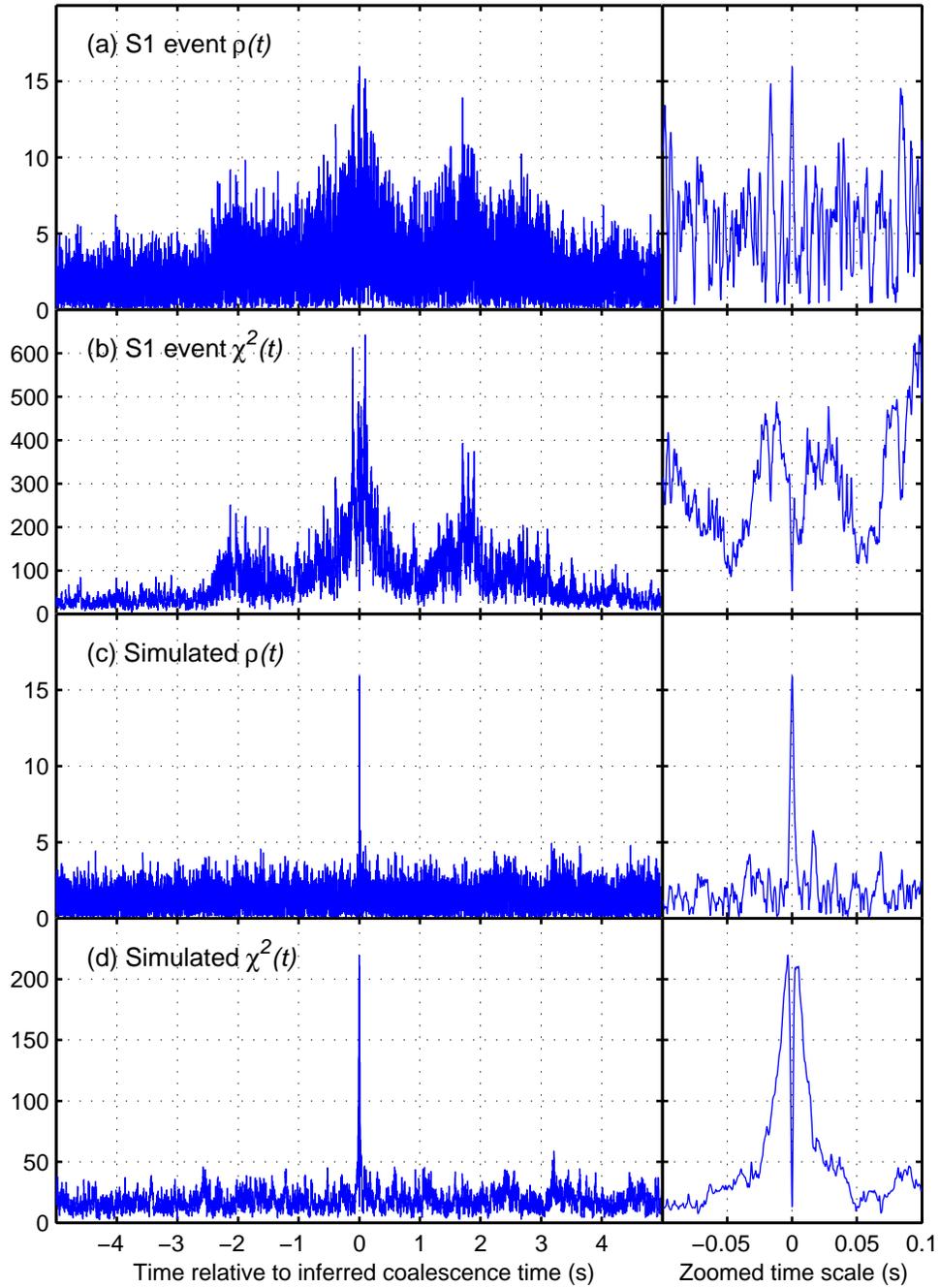}
  \end{center}
  \caption{Time series of ({\it a}) $\rho(t)$ and ({\it b})
  $\chi^2(t)$ in the vicinity of the highest-amplitude inspiral event
  candidate found by the LIGO S1 search, on two different time scales.
  For comparison, ({\it c}) and ({\it d}) show the time
  series for a simulated inspiral signal added to fairly well-behaved
  detector noise from a different time during the S1 run.}
  \vspace*{-0.3in}
  \label{fig:s1event}
\end{figure}

The search for binary neutron star inspirals in LIGO S1
data~\cite{S1BNS} did not find any {\em coincident} ``event
candidates'' with consistent signals in the two interferometers used
in the analysis.  Triggers from the {\em individual} interferometers
(when a coincidence check was not possible) which exceeded the
$\rho^\ast$ threshold and passed the $\chi^2$ test were
considered event candidates for purposes of calculating an upper limit
on the rate of inspirals in the Galaxy.

Follow-up examination of several event candidates with the largest
SNRs revealed that they did not resemble true inspiral events.  Plots
({\it a}) and ({\it b}) of figure~\ref{fig:s1event} show the $\rho(t)$
and $\chi^2(t)$ time series for the event with the largest SNR,
$15.9$, which was also shown in figure~5 of reference~\cite{S1BNS}.
Both time series are larger on average, and more variable, for a
period of several seconds around the inferred coalescence time
reported by the search algorithm.  (For comparison, the time series
expected for a real inspiral signal in stationary detector noise are
shown in figure~\ref{fig:s1event} ({\it c}) and ({\it d}).)  The
trigger was generated when the $\chi^2$ happened to fluctuate down to
a value below the threshold, which was 78 for this event according to
equation~(\ref{eq:modchisq}).  This particular event was found to have
been due to a saturation of the photodiode which produces the data
channel that was analyzed, but other large-SNR event candidates
without obvious instrumental causes show similarly anomalous behavior
in the time series.  Nevertheless, the search algorithm used in the S1
inspiral analysis found this event candidate, and others, because it
considered the values of $\rho$ and $\chi^2$ {\em only at a single
point in time}.

\section{Tests based on filter output history}
\label{tests}

Based on examining event candidates like the one shown in the previous
section, we concluded that a test based on the filter output {\em over
a time interval}, rather than just at a single point in time, should
provide an effective way to reject events like these which are caused
by non-stationary noise.  In essence, we wanted to find a quantitative
measure of the visually obvious difference between the data event and
the simulated event in figure~\ref{fig:s1event}, some sort of check
that the detector noise around the time of the trigger was consistent
with the stationary noise assumed by the matched filter.  Various
approaches are possible; we decided to focus on simple tests using the
$\rho(t)$ time series over a short interval just before the inferred
coalescence time.  We implemented a few potential tests by modifying
the ``findchirp''~\cite{findchirp} inspiral search code
in the LIGO/LSC Algorithm Library (LAL)~\cite{LAL}
and studied the effectiveness of these tests on the event candidates
found in the S1 data as well as on simulated events.  Note that the
computational cost of these tests is essentially zero, since the time
series of the filter output is already available in memory.

\begin{figure}[bt]
  \vspace*{0.2in}
  \begin{center}
  \includegraphics[width=5.0in]{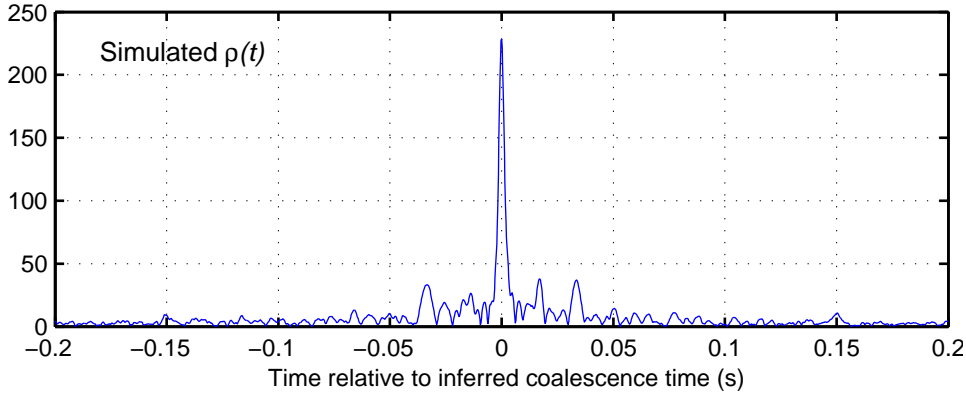}
  \end{center}
  \caption{Time series of $\rho(t)$ for a simulated inspiral signal
  with very large amplitude.  Detector noise, while present, is
  small on the scale shown.  The ``bumps'' on either side of the main
  peak are due to real correlations between the matched filter and
  time-shifted signal waveforms.}
  \label{fig:zoomsim}
\end{figure}

We found that a good way to distinguish real inspiral signals from
these triggers caused by non-stationarity is to count the number of
``crossings'', $N_c$, in the half-second interval leading up to the
inferred coalescence time, where a ``crossing'' is an instance of the
$\rho(t)$ time series crossing over a threshold value, $\rho^\times$,
in the up-going direction.  (Every trigger has at least one crossing,
as $\rho(t)$ rises to its peak value.)  We initially chose $\rho^\times
= 6.5$, which yielded large values of $N_c$ for many of the large-SNR
event candidates in the S1 data.
Small-amplitude simulated events, like the one in
figure~\ref{fig:s1event}({\it c}), had just one or occasionally two crossings.
However, we found that larger-amplitude simulated events often crossed
this threshold several times.  Figure~\ref{fig:zoomsim}
reveals the reason, showing the filter output for a simulated event
with very large amplitude: even when filtered with an exactly matching
template, the main $\rho(t)$ peak is accompanied by several additional
``bumps'' which far exceed $6.5$.  These bumps are due to the
autocorrelation of the waveform when time-shifted, and their exact
shape will depend on $S_n(f)$ since it affects the filter
(equation~(\ref{e:xfilter})).  For the S1 data, we found that the height
of these bumps relative to the peak was reasonably consistent for
inspiral waveforms with various parameters, so we settled on a
$\rho_{\rm max}$-dependent threshold of the form
\begin{equation}
  \rho^\times = \sqrt{ (6.5)^2 + (\rho_{\rm max}/6)^2 } \; .
  \label{eq:varthresh}
\end{equation}

\begin{figure}[tb]
  \begin{center}
  \includegraphics[width=5.0in]{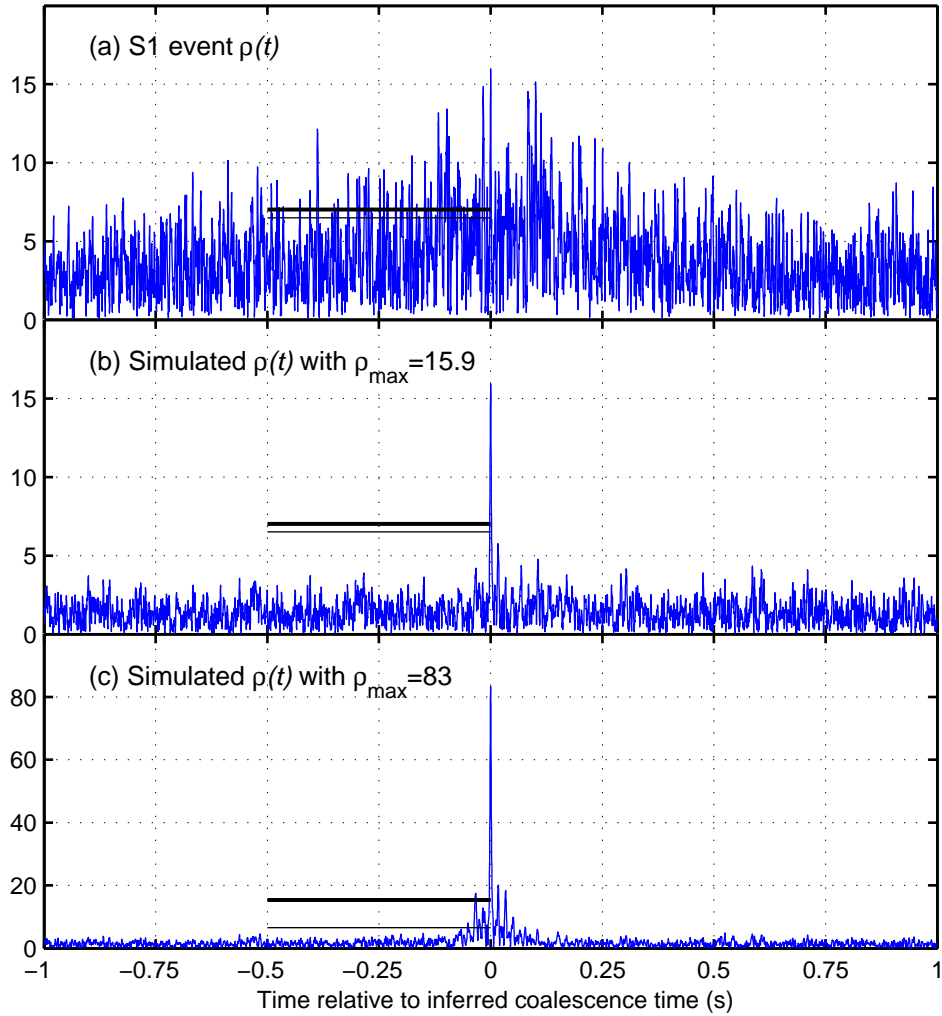}
  \end{center}
  \caption{Time series of $\rho(t)$ for: ({\it a}) the
  highest-amplitude inspiral event candidate found by the LIGO S1
  search; ({\it b}) a simulated inspiral signal with the same value of
  $\rho_{\rm max}$; and ({\it c}) a simulated inspiral signal with a
  much larger $\rho_{\rm max}$.  The two horizontal line segments in
  each plot indicates the half-second time interval for counting
  ``crossings'' with a fixed threshold of $\rho^\times=6.5$ (lower,
  thinner line in each plot) and with the $\rho_{\rm max}$-dependent
  threshold calculated according to equation~(\ref{eq:varthresh})
  (upper, thicker line in each plot).}
  \label{fig:thresh}
\end{figure}

Figure~\ref{fig:thresh}
illustrates the application of this test to the loudest S1 event
candidate as well as simulated inspiral events with small and large
$\rho_{\rm max}$.  The fixed threshold (thinner horizontal line in
each plot) and the $\rho_{\rm max}$-dependent threshold (thicker
horizontal line) are nearly the same for triggers
with $\rho_{\rm max} < 16$.  For large-amplitude simulated events, the
$\rho_{\rm max}$-dependent threshold is substantially higher and
avoids most of the bumps in the $\rho(t)$ time series; $N_c$ is
usually 1 or 2 for these simulated events.

\section{Application to LIGO S1 data}
\label{s1rerun}

\begin{figure}[tb]
  \begin{center}
  \includegraphics[width=5.0in]{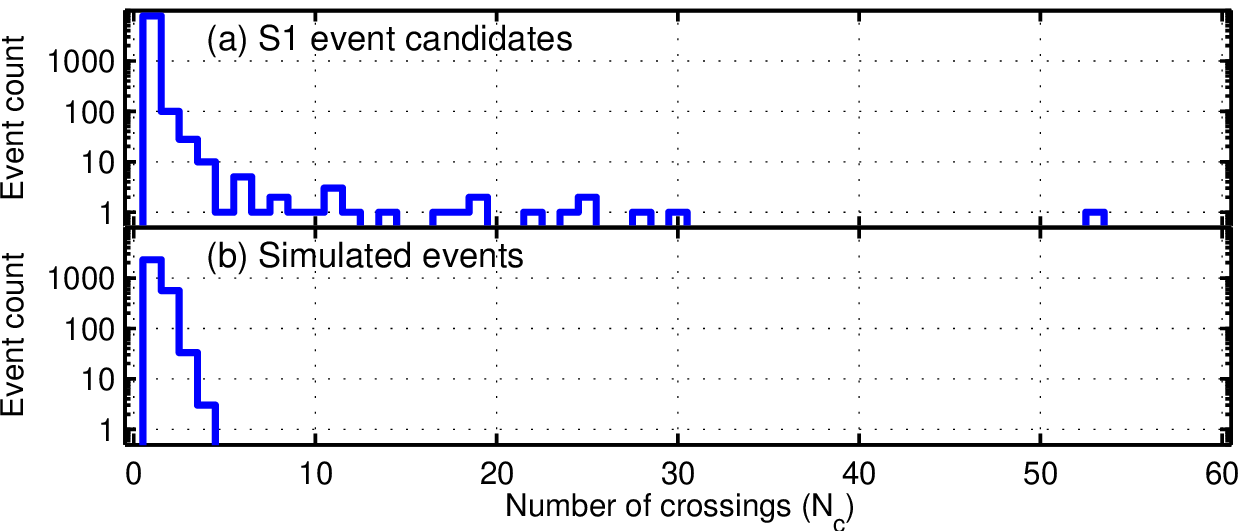}
  \end{center}
  \caption{Histograms of number of crossings for ({\it a}) data events
  found in the LIGO S1 search and ({\it b}) simulated inspiral signals
  from the population model used in that analysis~\cite{S1BNS}.}
  \label{fig:histos}
\end{figure}

\begin{figure}[tb]
  \vspace*{0.2in}
  \begin{center}
  \includegraphics[width=5.0in]{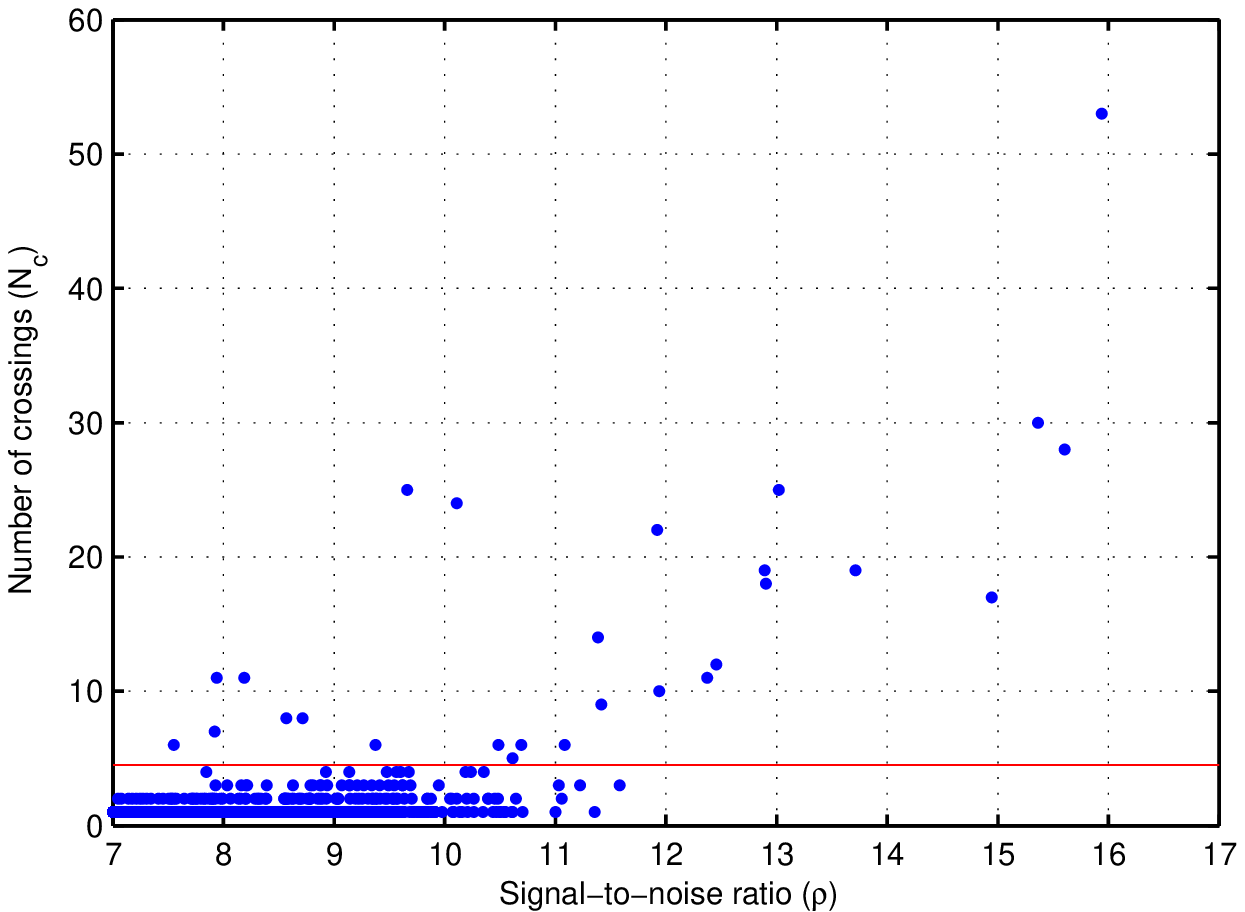}
  \end{center}
  \caption{Number of crossings vs.\ signal-to-noise ratio for data
  events found in the LIGO S1 search.  The horizontal line indicates
  the requirement $N_c \le 4$, which eliminates the 12 events with the
  largest values of SNR, as well as some events with smaller values of
  SNR.}
  \label{fig:scatter}
\end{figure}

We re-analyzed the full S1 data set, calculating $N_c$ for each event
candidate as described in the previous section.  We also re-analyzed
the set of simulated inspiral events generated from a population model
of the Galaxy and Magellanic clouds, which was used to evaluate the
efficiency of the search algorithm in the original analysis.
Histograms of $N_c$ for both samples are shown in
figure~\ref{fig:histos}.  None of the simulated events had more than
four crossings,\footnote{The fact that some simulated events had as
many as three or four crossings is thought to be related to
fluctuations in $\rho(t)$ when the contribution from detector noise is
comparable to the height of the bumps caused by the inspiral signal.
This could perhaps have been avoided by using a threshold with a
different $\rho_{\rm max}$ dependence.}
while a small but significant fraction of the data events had more
than four.  Figure~\ref{fig:scatter} shows that the event candidates
with the highest numbers of crossings are generally the ones with the
largest values of SNR, which are the ones we particularly want to be
able to reject.  In fact, by requiring $N_c \le 4$, we eliminate the
12 event candidates with the largest values of SNR.

This test was developed after the LIGO S1 analysis was finalized, so
it is not reflected in the published result.  In fact, the test was
tuned specifically to eliminate the large-SNR event candidates in the
S1 data sample, so it would, in principle, bias an upper limit
analysis based on that data sample.  Nevertheless, the ability of the
test to cleanly distinguish between real and false inspiral events is
very clear.

\clearpage
\section{Summary and Discussion}
\label{summary}

We have proposed a new type of waveform consistency test for binary
inspiral searches which uses the time history of the matched filter
output and which is complementary to the standard $\chi^2$ test.  A
simple test of this type, tuned heuristically using the LIGO S1 data,
was highly successful at eliminating large-SNR event candidates
without introducing any measurable inefficiency for real inspiral
signals.  We believe that this technique will be valuable for future
inspiral searches, although it will have to be tuned based on the
nature of the non-stationarity in the detector noise.  Within the
context of the number-of-crossings test, a different time interval
could be used, or the threshold could be chosen differently; it could
even have some functional dependence on time relative to the inferred
coalescence time.  There are, of course, many alternative ways to
derive a scalar statistic from the $\rho(t)$ and/or $\chi^2(t)$ time
series, with various threshold or distribution tests.  For instance,
Guidi has proposed using the maximum value of the filter output time
series, modified by subtracting the contribution expected from the
putative signal, over a time interval before the inferred coalescence
time~\cite{Guidi}.  The goal of any test of this type is to evaluate
whether the detector noise in the vicinity of the trigger is
consistent with the stationary noise assumed by the matched filter.

\ack

We thank Alan Weinstein for useful suggestions and Duncan Brown for
helping us to re-run the inspiral search on the LIGO S1 data.
This work was supported by the National Science Foundation through
Cooperative Agreement PHY-0107417 and through the Research Experiences
for Undergraduates (REU) program.



\Bibliography{99}

\bibitem{MF1} Cutler C and Flanagan \'E E 1994 \PR D {\bf 49} 2658--97

\bibitem{MF2} Balasubramanian R \etal 1996 \PR D {\bf 53} 3033--55
\par\item[] Balasubramanian R \etal 1996 \PR D {\bf 54} 1860 (erratum)

\bibitem{BCV} Buonanno A \etal 2003 \PR D {\bf 67} 024016
\par\item[] Buonanno A \etal 2003 \PR D {\bf 67} 104025

\bibitem{S1DET} Abbott B \etal 2004 \NIM A {\bf 517} 154--79

\bibitem{S1BNS} Abbott B \etal 2004 \PR D in press
\par\item[] (Abbott B \etal 2003 {\it Preprint} gr-qc/0308069)

\bibitem{SPA} Droz S \etal 1999 \PR D {\bf 59} 124016

\bibitem{GRASP} Allen B 2000 {\it GRASP: a data analysis package for
  gravitational wave detection, version 1.9.8 manual} pp 180--8
\par\item[] ({\it URL} {\tt
  http://www.lsc-group.phys.uwm.edu/\~{}ballen/grasp-distribution})

\bibitem{40m} Allen B \etal 1999 \PRL {\bf 83} 1498--1501

\bibitem{TAMAinspiral} Tagoshi H \etal 2001 \PR D {\bf 63} 062001

\bibitem{findchirp} Allen B \etal in preparation

\bibitem{LAL} {\it URL} {\tt http://www.lsc-group.phys.uwm.edu/lal}

\bibitem{Guidi} Guidi G 2004 \CQG submitted this issue

\endbib
\end{document}